\documentclass[a4paper,floatfix,rmp,twocolumn,showkeys,superscriptaddress]{revtex4}
\usepackage[latin1]{inputenc}
\usepackage{graphicx}

\begin{document}

\title{Rise of the humanbot}

\providecommand{\ICREA}{ICREA-Complex Systems  Lab, Universitat Pompeu
  Fabra,   Dr    Aiguader   88,   08003   Barcelona,   Spain}
\providecommand{\IBE}{Institut de Biologia Evolutiva, CSIC-UPF, 
Pg Maritim de la Barceloneta 37, 08003 Barcelona, Spain}
\providecommand{\SFI}{Santa Fe  Institute, 1399 Hyde  Park Road, Santa
  Fe  NM   87501,  USA}

\author{Ricard Sol\'e}
\affiliation{\ICREA}
\affiliation{\IBE}
\affiliation{\SFI}

\begin{abstract}
The accelerated path of technological development, particularly at the interface between 
hardware and biology has been suggested as evidence for future major technological 
breakthroughs associated to our potential to overcome biological constraints. This includes the 
potential of becoming immortal, having expanded cognitive capacities thanks to 
hardware implants or the creation of intelligent machines. Here I argue that several relevant 
evolutionary and structural constraints might prevent achieving most (if not all) these 
innovations. Instead, the coming future will bring novelties that will challenge many other 
aspects of our life and that can be seen as other feasible singularities.  One particularly 
important one has to do with the evolving interactions between humans and non-intelligent 
robots capable of learning and communication. Here I argue that a long term interaction can lead to a 
new class of ``agent'' (the {\em humanbot}). The way shared memories get tangled over time will inevitably 
have important consequences for both sides of the pair, whose identity as separated entities might 
become blurred and ultimately vanish. Understanding such hybrid systems requires a second-order 
neuroscience approach while posing serious conceptual challenges, including the definition of consciousness.
\end{abstract}


\maketitle

\begin{quote}
\begin{flushright}
{\em Can a machine think? Could it have pain?}\\
Ludwing Wittgenstein
\end{flushright}
\end{quote}

\section{Introduction}

The beginnings of the 21-st century have been marked by 
a rapid increase in our understanding of brain organisation and a parallel improvement 
of robots as embodied cognitive agents (Steels 2003; Cangelosi 2010; 
Nolfi and Mirolli 2009; Vershure et al 2014). This has taken place along with 
the development of enormously powerful connectionist systems, particularly 
within the domain of convolutional neural networks (Lecun et al 2015; Kock 2015). Two hundred years 
after the rise of mechanical automata, that became the technological marvels of the Enlightnment (Woods 2003)  
new kinds of automata are emerging, capable of interacting with humans in adaptive ways. 
The requirements for building an intelligent or a conscious machine are likely  to be still ahead 
in the future, but some advances and new perceptions of the problem are placing the possibility 
at the forefront of "what-if" questions (Vershure 2016). To a large extent, today's discussion on 
what separates humans from their artificial counterparts is deeply tied to the 
problem of how to properly define mind and consciousness (Zarkadakis 2015).

\begin{figure*}
{\centering 
\includegraphics[width=17 cm]{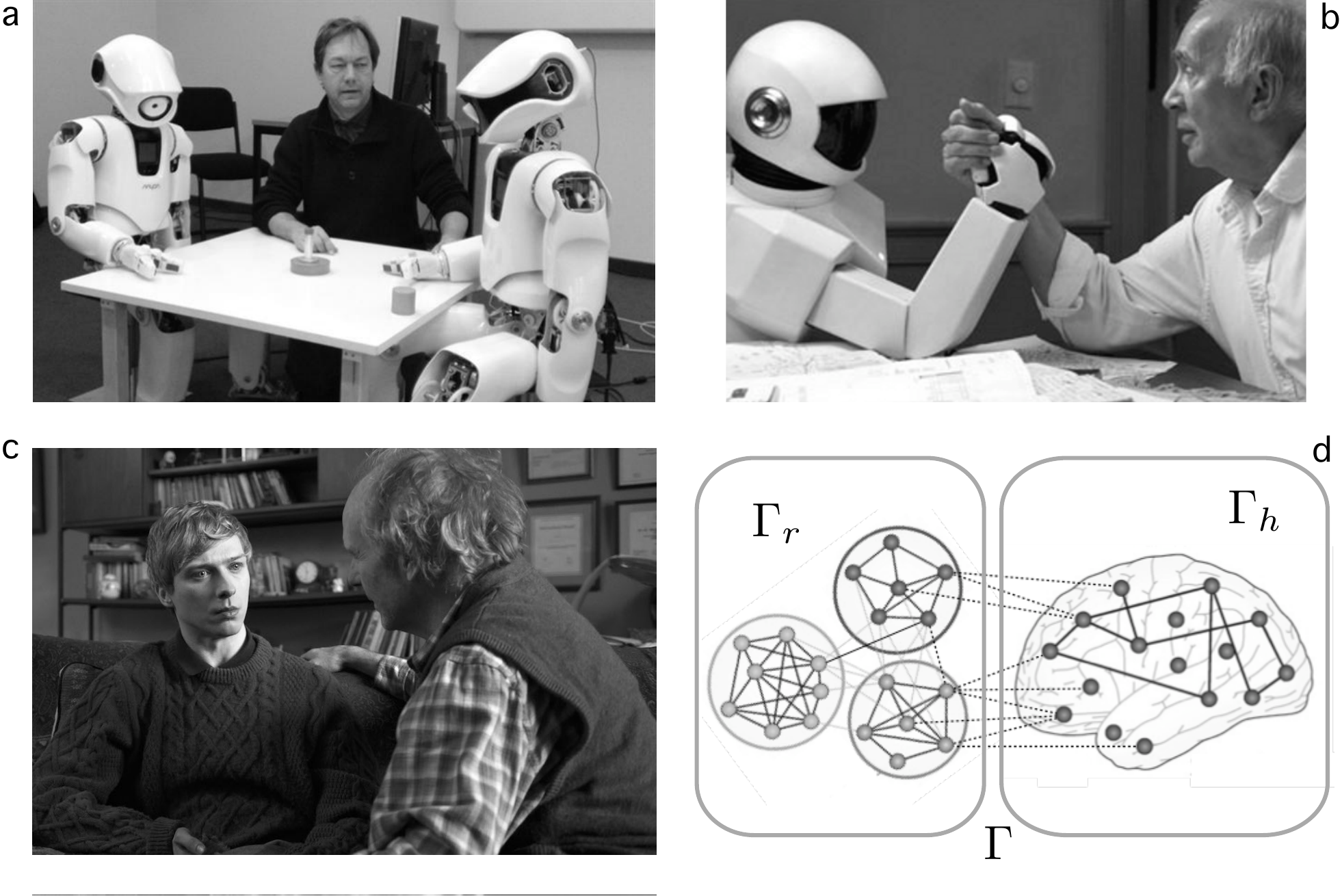}
\caption{Humans and robots have been interacting 
 in increasingly more complex ways, most of them limited to simple tasks. However, 
 communicating robots (a) could in the future interact in deeper ways both among them and 
 with humans (image courtesy 
 of Luc Steels -the human in the picture- and the Neurocybernetics group at Osnabruck). Fictional stories 
 on future human-robot interactions, including (b) the movie {\em Robot and Frank} or (c) TV series {\em Humans} 
 have started to consider the relevance of strong emotional ties established between elderly or 
 impaired human beings and (possibly non-intelligent) robots capable of learning and 
 communicating through natural language. Evolving connectomes will inevitably result 
 from HRI (d). The long-term association between a human and a robotic agent, when the 
later is equipped with both communication skills and learning capacities implies the 
creation of an association network $\Gamma$ linking both "brains" through a number of 
shared memories. The relative importance of this shared connectome will be a function of 
the cognitive apparatus of each partner and the depth of emotional engagement. 
}
}
\end{figure*}

In the 1950s, the development of cybernetics by Norbert Wiener and others
 along with the beginning of theoretical neuroscience 
and Turing's proposal for an intelligence test  (Turing 1950) were received with a 
similar interest, triggering 
a philosophical debate on the limits and potential of man-made imitations of life (Nourbakhsh 2013). The study 
of the first "cybernetic machines" made by a few pioneers such as Gray Walter generated 
a great expectation. For the first time "behaviour" emerged as a word associated to mechanical 
machines, this time empowered by the rising technology that allowed to combine hardware and 
a new form of engineering inspired -to some extent- by natural devices (Water 1950, 1951; 
see also Braitenberg 1984). Those 
early experiments provided some interesting insights into the patterns of exploration 
of simple agents that where able to detect edges, respond to light and modify their movements 
according to some simple feedback mechanisms.  Although their simple behaviour was essentially 
predictable, it was not completely free from surprise (Holland 1997). Later work in the 1990s and afterwards 
incorporated artificial neural systems as an explicit form of introducing learning and behaviour 
largely inspired in biology (Vershure et al 1992; Edelman 1992; Sporns and Alexander, 2002; Prescott et al 2006).

Nowadays we are 
rapidly moving towards a new generation of domestic, human-friendly robots that will probably trigger 
a new technological revolution, similar in some ways to the one proposed by the first 
personal computers. There is a great promise in this revolution. The promise of 
robots helping us, playing with us or simply having some basic support functions is pushing 
forward software and hardware companies towards designing cheap robotic systems. 
For the general public the fascination remains in the humanoids, those who 
really look like us. We are actually witnessing a rehearsal of the mechanical automata 
mania of the 18th century, except that the new automata will be much more autonomous 
and perhaps closer to us than ever. Not surprisingly, the rise of the new robots has come about with 
a parallel growth of new fields associated to human-robot interactions 
(Breazeal 2003; Fong et al 2003; 
Dautenhahn 2007) and the exploration 
of learning, cognition and evolution potentials of artificial agents 
(Clark and Grush 1999; Schaal 1999; Floreano and Mattiusi 2008; Takeno 2013).

Along with the science and engineering of these human-like agents, science fiction 
is also exploring some relevant (and sometimes unexpected) consequences of 
these near futures. Most early works on robots have been centered on the 
possibility of intelligent, or even self-aware machines. Isaac Asimov published the 
Sci-Fi classic {\em I robot} three years before Turing's landmark paper on intelligence, picturing 
a would-be society where robots become 
a central part of our lives (Asimov 1947). Perhaps the most interesting reflection of Asimov's tales 
is the unexpected, the unforeseen consequences of robotic actions while interacting 
with their most complex part of the environment: us, the humans. The robots incorporate 
a set of hardwired rules (the "laws" of robotics) preventing the machines from harming 
humans and harming themselves but in special contexts, when a conflict among the laws 
emerges, the unexpected should be expected.  

The scientific progress made in the field of human-robot interactions (HRI) and artificial 
intelligence (AI) have renewed the interest in the potential consequences of advanced cognitive robots 
(Wallach and Allen, 2009; Bradshaw et al., 2004; Murphy and Woods, 2009) and 
in particular the role played by embodiment in communicating agents (figure 1a) which 
is known to play a key role in the development of complex behavioural traits (Steels 2003, 2015). 
The plots in new Sci-Fi stories have become more subtle, and much more interesting. 
In the movie {\em Robot and Frank}, for example, we have a 
moving story of a declining man (Frank) facing the initial phases of Alzheimer's and a robot companion 
which, despite a lack of real intelligence, shares experiences and memories with his human partner (fig 1b). At some point, the robot asks Frank to wipe out his memories. Frank rejects the idea, while the robot 
tells him "I am not a person, I am just an advanced simulation" but afterwards he uses a number of 
very personal (Frank-like) sentences to make his argument, resulting from previous learned 
talks and experiences. A related situation also 
shows up in an episode of the Sci-Fi series {\em Humans} where Dr George Millican, an elderly man 
and one of the main characters, owns a faulty robot named Odi (fig 1c). The boy-like robot is an old 
model that stores memory chunks of shared, unique experiences, some of them related to deceased Millican's wife. Because of this, Millican does not accept the mandatory replacement of old models, thus hiding Odi inside a closet. 

None of the previous examples is related to truly intelligent robots. Instead, the 
subtle ties and their consequences arise from shared cognition and emotions. 
Nowadays, despite our distance from the sci-fi proposals, 
robot designs are already present in a wide range of situations where they 
are social partners. This includes household pets, healthcare assistants or even educational 
companions. Even when the robotic agent has a limited repertoire of interactions, 
its actions can be perceived as part of some "personality", particularly in relation 
with robotic pets (Min Lee et al 2006; Li and Chignell 2010; Miklosi and Gacsi 2012; Park et al 2012). 
One goal of robotic research is the development of autonomous robots, capable of perception 
but also of making decisions by themselves and more importantly to communicate (Breazeal 2002; 2003; 
Breazeal et al 2016). 
Here several levels of complexity are possible, from programmed agents responding to a more or less 
simple and predictable environment to robots capable of complex interactions with humans, 
able to use natural language and learn from experience. The later scenario 
implies sophisticated implementations grounded in neurorobotics 
(Arbib et al 2006; Oztop et al 2006; Arbib et al 2008). One goal of this paper is actually 
provide a general space of HRI complexity where different classes of interactions can be 
located. 

As will be argued in the next section, these shared memories and other common relationships 
define a network of interactions describing the HRI that can contribute, 
but also replace, parts of the brain functions (figure 1d). This diagram displays an idealised 
bipartite system ($\Gamma$) including both the human cognitive network ($\Gamma_h$, the 
brain) and the artificial neural network embedded in the robotic agent ($\Gamma_r$). The 
whole network $\Gamma$ will change in time as changes in the two subnetworks occur 
but  also through the creation of all levels of correlations emerging from the HRI. 
What happens when this network 
is the outcome of a long-term exchange? What are the consequences for a HR pair 
when the human node is affected by some kind of impairment? As will be argued below, 
by considering the relative cognitive complexities of each agent in this 
HRI and the influence of the degree of emotional engagement, a space of HRI can be defined following 
the approach of second-person neuroscience (Schlibach et al 2013). 
Within this space of possibilities, one in particular can involve the emergence of a new class of cognitive 
agent transcending both humans and robots, perhaps 
defining a novel form of synthetic evolutionary transition (Sol\'e 2016). The consequences of such novel hybrid system will be discussed.

\section{The space of human-robot interactions}

Here we are interested in the problem of how a socially interacting robot can 
trigger emotional and even cognitive changes in the human partner. In order to 
do so we need to take into account several components of this interaction, which is mediated by 
a complex network of exchanges and could be described 
in the context of distributed adaptive control theory (see Lallee and 
Vershure 2015). Two major groups of HRI interactions can be defined here. This first involves those scenarios 
where interactions are short-lived (the 
robot companion exchanges take place over a small time window) or predictable (the 
robot is programmed for simple tasks). The second instead incorporates robotic agents 
that can learn from experience and interact through long time spans. A major difference 
between the two cases, in terms of the HRI is the absence and presence of emotional 
engagement (EE). The choice of EE as a key axis in our approach is grounded in previous 
studies on second-person neuroscience (Schlibach et al 2013). In this field, the key assumption 
is that an active social engagement between two (or more) individuals is fundamentally 
different from a mere observation of subjects. Within our context, we will distinguish 
between a situation of low EE where the two partners in a HRI are essentially independent 
from another one where each agent shapes others cognition.  

The space of HRI proposed here is summarised in figure 2a, where three axes have been 
used to locate different classes of robotic agents. In a way, this 
can be seen as the landscape of human-robot interactions. It provides a tentative picture of 
the diverse array of qualitative classes of HRI. Three axes are introduced, namely: the degree of 
cognitive complexity of robots and humans and a third axis linked to the emotional engagement, 
resulting from both a long-term interaction and the potential for learning and adaptation displayed 
by the robot. As shown in figure 2b and 2c, we can decompose this space, which we have partitioned 
in eight arbitrary combinations, in two main layers including low and high 
emotional ties. Below we present and discuss these two layers separately.

\subsection{Low emotional engagement}

In this level of engagement, robots are typically associated to simple tasks with 
low (if any) of cognitive complexity and playing the role of decision-making 
systems exhibiting low communication skills (figure 3b). Most standard socially interacting robots 
(at least the first generations of them) fall in this domain. The typical scenario is 
a programmed artificial agent that can perform predictable tasks and is expected 
to operate in simple environments. 

Humans (from toddlers to 
adults) interacting with simple robots (such as toys or automata following simple orders)
 would define a lower bound of low cognitive complexity for both agents. Cleaning robots 
 such as Roomba and receptionist robots answering requests from clients (giving 
 simple types of information) would be obvious examples, although even in this case 
 the use of personalization toolkits triggers emotional connections (Sung et al 2009).
 
In this domain, robots with high cognitive skills interacting with cognitively impaired individuals 
over specific tasks give the last option in our list. Robots can help in providing 
support to the elderly or impaired in ways that do not require learning potential from the 
artificial agent side. Robots capable of identifying the needs of their human partners 
can be helpful even if not emotionally engaged, while robots with a large amount of 
available data sets can be interesting as expert systems with user-friendly (humanoid) interfaces. 
Instead we weight cognition in terms of (pre-programmed) diverse repertoires of responses.
 
Agents and operative systems exhibiting a rich repertoire of interactions, such
 as SIRI (which can assist blind people) or Alexa would fit the high-cognitive complexity corner. 
 Here of course "cognition" does not 
 have the meaning that can be attributed to a neural system. For some of these non-embodied systems, 
 a very large repertoire of potential answers can be communicated with the help 
 of a natural language interface. The success of some systems such as Watson, which was 
 trained for a specific goal (answering questions on the Jeopardy quiz show) using 
 hundreds of millions of webpages illustrate the potential for surpassing humans in searching 
 and solving questions.

\begin{figure*}
{\centering 
\includegraphics[width=11 cm]{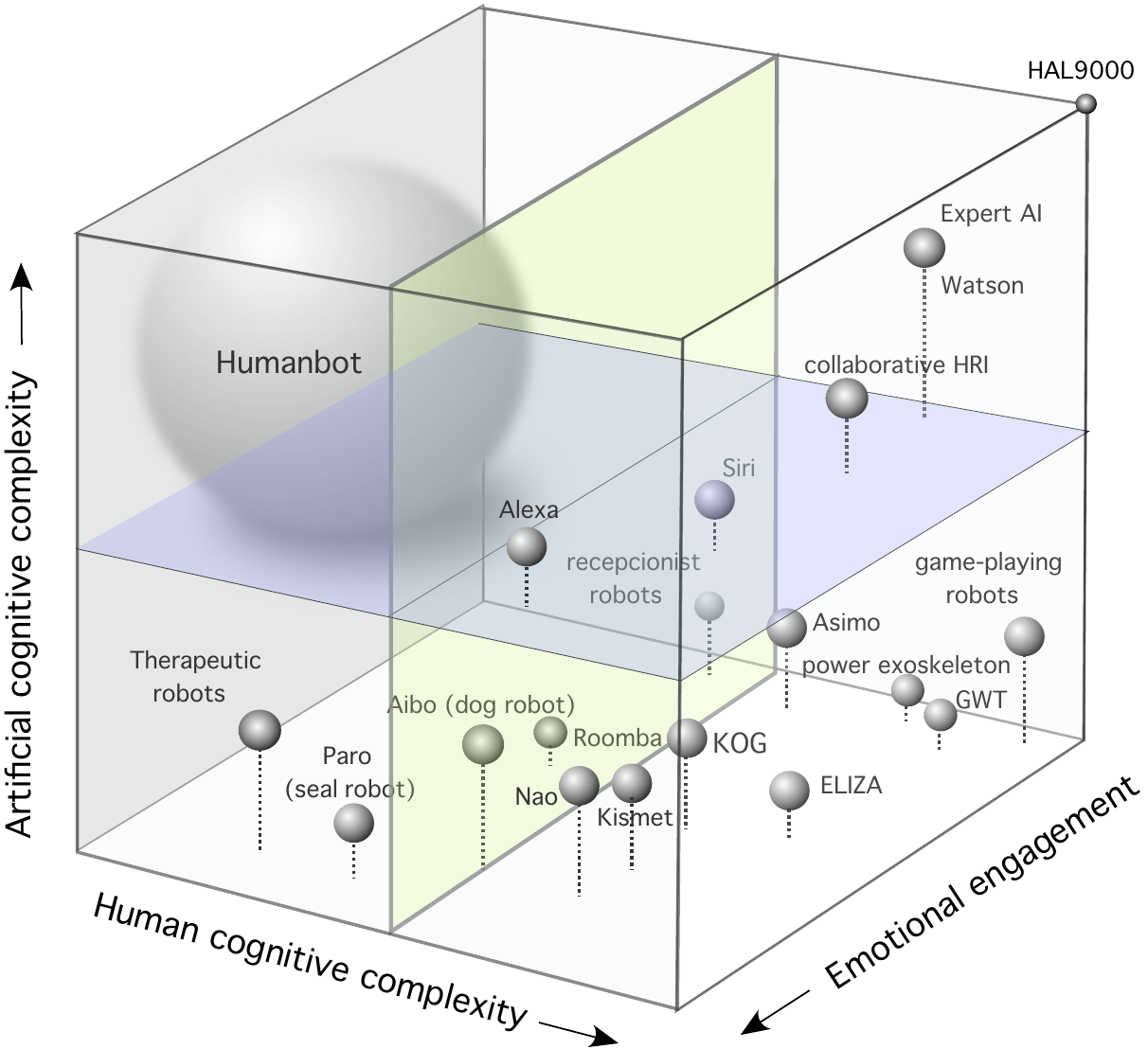}
\caption{The cognitive space of H-R interactions under the two-agent perspective taken here. Non-robotic 
systems (such as ELIZA, SIRI or Watson) are also included. The 
axes represent (in a qualitative ordering) emotional engagement as well as two dimensions associated to the 
complexity displayed by the human and the artificial agents, respectively. Here the relative location 
of each system needs to be taken as indicative. Some of these robots appear in the high human 
cognitive complexity domain, mainly because their full operational function is expected to take place 
here, although they could in principle be useful in the other side of the cognition space. The domain 
where {\em humanbot} systems would emerge is indicated by a blurred sphere. A broad array 
of possible HRI pairs could be expected here, involving strong cognitive 
dependences that could generate, particularly when the human side is impaired, a new class 
of cognitive agent.}
}
\end{figure*}

 This layer and the next layer in our space, now allowing higher emotional engagement, are not 
 clear cut. This is particularly relevant if we take into account the human tendency to extract 
 behavioural or emotional clues from the interaction with even simple computer programs. 
 It is worth remembering that even the earliest attempts to program machines (Weizenbaum 1966) capable for 
 answering questions, such as ELIZA (which used pattern matching to simulate a conversation) led 
 to rather unexpected reactions (Weizenbaum 1976). ELIZA was supposed to imitate a speaking 
 psychiatrist, essentially creating simple responses triggered by key words that were 
 then used to create simple questions. But in many cases, people failed to believe 
 they were talking with a computer (which was far beyond 
 in computer power from anything existing today). Similarly, some robots can be simple companions if their 
 interactions span a short time scale but create a strong bond (from human to machine) with their owners 
 if interactions occur over long, shared periods of time. To reach the second layer, we 
 must allow artificial agents to learn and adapt in flexible ways, as well as be capable of exhibiting and 
 detecting emotions.

\begin{figure*}
{\centering 
\includegraphics[width=14.5 cm]{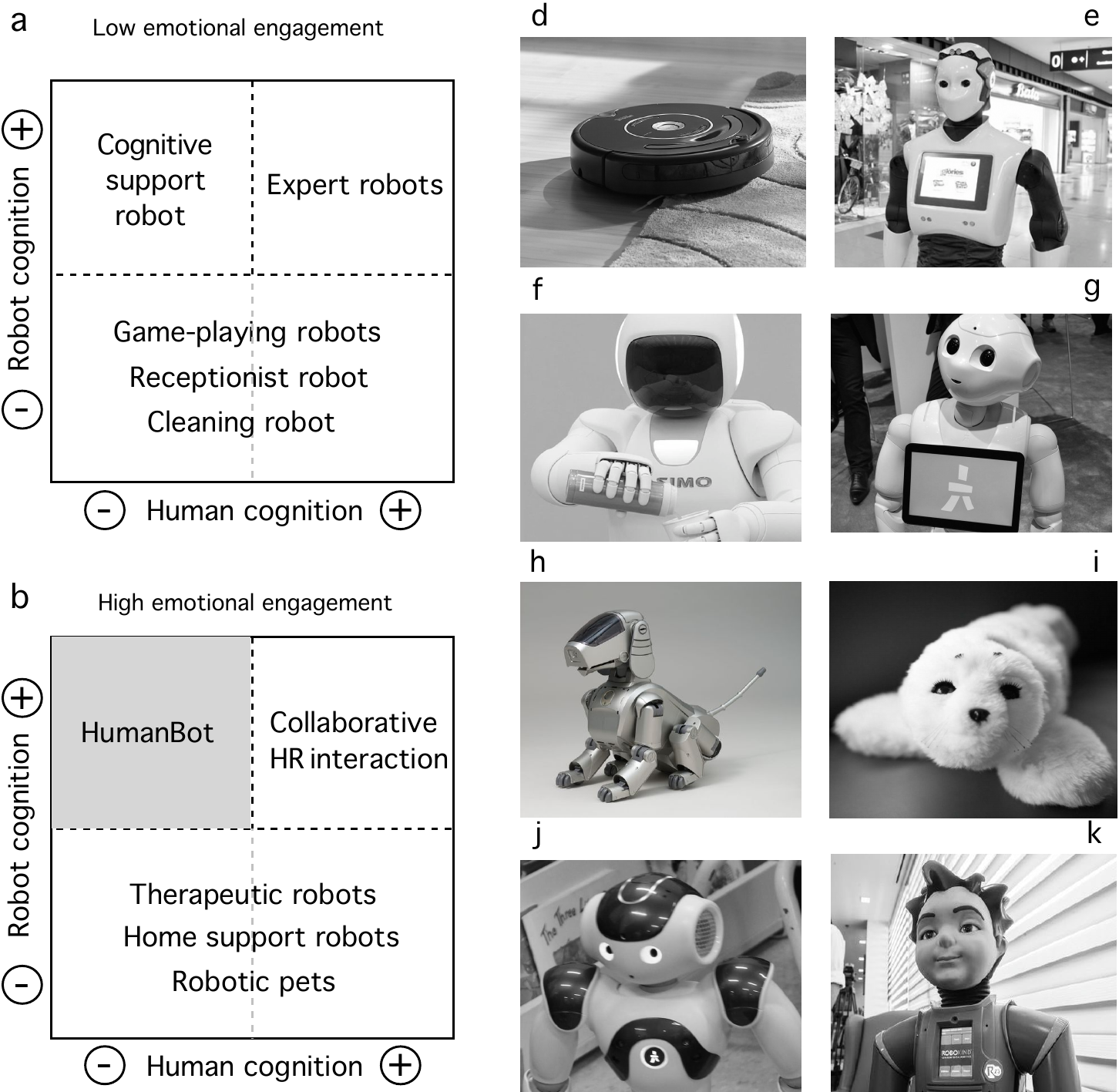}
\caption{The cognitive space of H-R interactions (a-b) here the two layers associated to different 
emotional engagement (related to short/long term interactions) have been 
separated indicating potential examples within each discrete class. Several examples of 
robotic agents engaged in HRI are shown in d-k, including programmed systems with little or no emotional 
responses nor learning capabilities to robots capable of simulating emotions, seeking eye contact and other 
features that enhance human emotional and social responses. Here we show: (d) Roomba cleaning robot, 
(e) REEM, from PAL robotics, (f) ASIMO, (g) Pepper, (h) Aibo, (i)  Paro the seal, (j) Nao and (k) Milo.}
}
\end{figure*}

\subsection{High emotional engagement}

Long-term relationships in HRI can lead to a rather distinct set of patterns that strongly depart from the 
previous layer. Here we consider the possibility that the robotic agents have been interacting 
with a given individual in a flexible manner and over an extended period of time. 
Such interaction might occur at different levels and this too is strongly dependent on the 
relative cognitive complexity or each partner (the human and the robot) as well as the  
level and span of their emotional interaction. A healthy, adult human brain is a highly 
complex system that can nevertheless engage in an emotionally strong relationship with 
a pet. Humans and dogs (and other pets) have been co-evolving  over 
hundreds of thousands of years until today, where our animal companions have limited cognition powers 
but a highly developed set of skills associated to emotion recognition (Hare et al 2002). One 
side effect of this process is our tendency to generate empathy for non-living objects resembling pets. 
These evolutionary responses have left a detectable cognitive signal (Stoeckel et al 2014). 
What has been the outcome of HRI between human and artificial pets? A first glimpse of 
the implications of long-term exchanges between humans and pet robots capable of learning 
was provided by AIBO dogs, used as companions in a broad range of conditions, from 
preschool children to elder patients (Kahn et al 2006, Melson et al 2009). In the
future, these type of pets (as well as those emerging from the Virtual reality world) are likely to become 
the rule rather than the exception (Rault 2015). 

Despite the lack of a complex communication, the domestication of dog's cognition has lead to 
emotional ties that  can be strong. All studies on HRI involving robotic pets reveal 
that human perception is markedly biased towards perceiving them as life-like 
entities and treated as such, instead as artifacts (Kahn et al 2006). This creates a number of interesting situations 
relevant for the problem addressed here, an in particular the blurring of lines between ontological 
categories. As a consequence, emotional ties and their consequent effects are 
likely to be shared. Such effects are enhanced by the development of some "personality" 
associated to robot learning capacities, which can strengthen emotional ties. These differential 
behaviours result from the historical sequence of HRI events: by tapping 
AIBO's head sensor after a given behavioural display, it is possible to enforce or decrease a given 
response, but each case and how it relates with other responses will depend on each specific HRI. 
Among other things, the loss of a pet can trigger 
a strong reaction in the human, but also in the animal end of the tie. Perhaps not surprisingly a similar situation has been emerging in relation with AIBO owners who, unable to repair their old dog robots, 
have to accept their "death" and eventually organise funerals not different from those 
made for the living counterparts. In this case, the human and the robot are separated by 
a huge cognitive complexity gap, but the capacity of the robot for learning and somewhat 
adopt some personality makes a big difference for their owners. 

The previous example would correspond to the possible interactions in the 
lower part of our space (figure 2c) associated with 
low robotic cognition. A relevant scenario is the interaction between elderly patients 
requiring care and therapy (the so called fourth age) and  some robots, such as Paro: 
an artificial, seal-shaped agent. Paro is equipped with sensors detecting touch, light, sounds 
or movement, with motors and actuators and other 
standard robotic features but also with some additional features such as responses to cuddling 
and a constant seeking of eye contact. Additionally, several emotions have been programmed 
and it can respond to its own name as well as learn other names. Paro has been used 
in treatment of patients with dementia. The result of these interactions was a reduction 
of anxiety and helping recover from chronic ailments as well as improving communication with 
other patients and often creating strong attachments (Kidd et al 2006, Takayanagi et al 2014). 
Similar patterns have been found using AIBO (Melson et al 2009). 

Most companion robots helping patients with dementia are programmed 
to respond to specific tasks in a more or less flexible way. On a basic level, the robot can be very useful by reminding the human the name and/or location of objects. This has been proposed as a memory prosthetics 
for elders (Ho et al 2013) where the robotic companion would be equipped with a visual 
episodic memory, but consider now a HRI  
capable of learning and using natural language.  On another level, it can serve 
as a medium to communicate with other humans and preventing isolation. Similarly, 
if capable of moving in outdoors environments, it could greatly improve orientation. 
But what if the artificial agent is equipped with a powerful 
cognitive system beyond simple reactions and capable of dealing with daily living environments 
using multimodal integration learning? Here deep neural networks can play a fundamental 
role, including the use of natural language (Noda et al 2014). 

Consider a robotic agent capable of maintaining a conversation, using natural language, 
and capable of gathering relevant information concerning past events related with the 
life of the patient, detecting goals and wishes as well as emotions and capable also 
of facial expression to share emotional states. This would also be a very desirable 
form of interaction, since the loss of memories or -more generally- the difficulties to access 
stored recollections can be a great source of stress for elder people with mild cognitive 
impairment\footnote{This refers usually to a transition stage between a normal process of brain ageing 
and dementia, characterised by low performance in memory tasks} and 
Alzheimer's patients in particular, specially in early 
phases of the disease. An artificial agent capable of coping with memory 
decay and disorientation by means of verbal communication would leverage the anxieties 
associated to cognitive decline and improve reasoning and judgement. If flexible enough, a neural agent 
can learn how to help the human in the most useful and personalised way. And here comes the problem.

The HRI involved here leads to a rather unique outcome. For example, if the artificial agent provides 
support to losses in episodic memory, a first paradoxical situation might emerge: it can occur that 
some events become absent from the impaired brain while they remain stored 
within the artificial agent. Such stored recollections can be easily over-interpreted by the 
artificial agent, in particular their relative value and potential correlations 
among different memories. On a general level, 
reminding the human subject where are the lost keys or a misplaced wallet are simple and yet 
important tasks that can easily counterbalance or alleviate early symptoms. But it can also 
have a major impact on autobiographical memory. This class of memory, associated to the left 
prefrontal cortex, provides the basis for putting together a timeline of past events connected to 
visual and sensory-perceptual features. We should have in mind that a major limitation of 
robotic agents (and a crucial component of the human mind, Suddendorf 2013) is their lack of 
understanding or representation of time. 

The agent, if properly equipped with visual recognition systems, 
can have seen pictures of family members (alive or not) and learned about their stories from the 
patient. These stories can be true or not, but in both cases the agent will contribute to store and 
recover them. A great advantage of any neural-based system capable of pattern recognition and 
generalisation (as deep learning networks, see LeCun et al 2015, and references therein) 
is that the robot can extract correlations required 
to help in more complex tasks, particularly in relation to episodic, semantic and working memory as well as 
language and executive functioning. But these correlations rely on both the (possibly faulty) input provided by the 
human and the emotional weight given to each memory by both partners within the HRI. Because correlations are likely to be generated from biased perceptions, the resulting internal correlation matrix created by the robot 
(and returned to the human through HRI events) can depart from the original correlations generated 
in the brain. 

If we take into account that memories themselves are not reliable (this in particular affects 
priming) the long-term HRI inevitably leads to considerable deviations from the original memory web. 
In this respect, we have a first glimpse of an anomalous pattern: potential memory deficits are 
compensated by the reliable storage of information residing in the artificial neural network, which 
can use false memories, create incorrect (but strong) correlations associated to emotional events and 
feed-back them into the mind of the human. Can this process lead to a runaway amplification phenomenon? 
What seems likely to occur is that mismatches between the relevance and emotional weight associated 
to different memory patterns and their interactions might promote deep distortions of the memory and 
behavioral landscapes.  As the HRI proceeds, a new set of interactions will coalesce between 
the human connectome $\Gamma_h$ (figure 1d) and the network of neural correlations created within the 
robot brain ($\Gamma_r$). The whole cognitive map must be found in the merged structure $\Gamma$ 
that includes both networks along with all the human-robot correlations that have emerged and that 
we also indicate as links between areas. 

The loss of plasticity that results from ageing or 
damage reinforces dependencies among human and robotic cognitive maps mediated by co-occurrence 
patterns. The view of a given object or image elicits a response in both sides that defines effectively 
an interaction between both, since these responses will immediately lead to an information exchange. 
Consider for example a picture of someone (or any other representation of it) 
that is identified by the human and also stored by the robotic companion. Previous exchanges 
will have weighted the relevance of this picture and the associated subnetwork of related objects, 
people or actions. 

As cognitive impairment grows in time, the relative importance of the object representation within 
the artificial system might have been enhanced beyond its original relevance, while exciting and reinforcing 
other related subnetworks. If decisions or actions derive from this perception, the loss of proper 
decision making by the human might have been displaced towards the robot, thus shifting the 
deep correlates from one agent to the other and helping to preserve episodic and semantic 
memory. Alternatively, different perception and decision layers 
might become segregated between them. If the memory of this specific image is gone from 
the human, it can nevertheless remain accessible in the artificial side, which would now contain part of the 
autobiographical memory. 

Since other related events connected to this memory might 
indirectly interact with the brain network, it is possible that novel forms of hybrid memory might 
emerge. However, despite the positive side of keeping otherwise erased memories, the 
dynamics associated to the formation of the HR network can easily shift the importance of 
events over time and even disrupt it. Unless under some external supervision by close relatives 
of the patient, the humanbot can lead to a shared mind that strongly departs from the 
organisation and coherence of the original subject. In other words, one outcome of this HRI 
is the conscious experience of a different subject, like living in the mind of someone else. 
The humanbot is a likely outcome of future HRI and its potential to become real is tied 
to the new generations of robots equipped with powerful learning systems and high memory 
capacities. These robots might be not yet here, but they are certainly much closer than the coming 
of intelligent machines.

\section{Discussion}

The increasing frequency of dementias, being Alzheimer's the most common one, will affect millions 
of human beings in the next decades (Reitz et al 2011). While prevention strategies are developed 
and new drugs are been tested, the need to caregivers helping these patients is becoming a major 
issue. Beyond the staggering economic costs, caregivers (often family members) are also 
affected by strong physical and emotional stress. In many cases, their health is also deteriorated. 
The possibility of using a robotic agent providing help seems a desirable option, although ethical issues 
need to be considered (Ienca et al 2016). Because cognitive 
decay is a central issue in most cases, the artificial agent should be equipped with a flexible, adaptive 
system capable of dealing with changing conditions and specific needs. But such plasticity 
and learning potential can generate new emergent phenomena. In this paper we have explored the potential outcomes of long-term HRI with artificial 
agents able to replace memory deficits and communicating through a natural language 
interface. As discussed above, the increasing replacement of faulty cognitive networks 
is likely to create a profound dependency of the human companion that can eventually 
end in a blurred boundary between the robot and the brain. 

There are other implications derived from this class of long-term HRI: 

\vspace{0.2 cm}
(a) The theory of attractor neural networks (Amari and Maginu 1988; Amit 1992; Rojas 2013) 
has shown that the qualitative responses 
of neural networks concerning their responses to noise, memory potential and other properties 
can experience sharp changes as some parameters are tuned. One particular example 
is the rapid decline of associative memory as the neural network is damaged beyond a given 
threshold. Similarly,  the qualitative changes associated to an mismatch between the 
memory requirements and the available cognitive power can lead to the emergence of spurious 
memory states that is decoupled from the real repertoire of original memories. In all these 
cases, neural networks display different {\em phases} separated by well defined phase transition 
points (Amit 1992). If these results, grounded in simple neural network models, can be extrapolated to 
our hybrid system, we would expect to observe tipping points in the cognitive organisation of 
memories and other key properties. 

\vspace{0.2 cm}
(b) human beings display awareness, while machines (so far) do not. A relevant 
problem concerning consciousness is how to define it and even how to measure it (Tononi and Edelman 1998). 
Some efforts in this direction suggest that a single parameter $\Phi$ could be 
defined that can capture the {\em degree} of consciousness (Tononi 2012). Using this 
parameter, obtained from an information theoretic approach, it has been argued that 
we can to the least order different case studies, from animals to impaired human brains or machines 
 (Tononi and Koch 2015). An interesting outcome of this approach is the suggestion that machines (in particular those based on von Neumann architecture) are not conscious (Tononi et al. 2016). Without discussing this conclusion, it seems clear that the humanbot, 
 by inhabiting the boundaries between human and machine will also incorporate some level 
 of consciousness (as measured by $\Phi$). Under the conditions described above, we need to ask the impact of 
 the cognitive replacement associated to the increasing interdependence and how is consciousness 
 shared by both parts. 
 
\vspace{0.2 cm}
(c) Once the human is gone, we should seriously evaluate what is left behind 
within the robot. As long term interactions are likely to shape the 
robotic cognitive network, some key components of the human's mind might 
remain there for inspection or preservation. Once deceased, what is left can keep 
changing as other inputs from the external world keep entering the system, thus 
modifying or even erasing the previous stored memories. What to do next? Should 
the capacity for inspecting the environment be put on hold? Would the robot be 
capable of interacting with friends and close people of the deceased in meaningful ways? 
Should the stored information be preserved as the only relevant remain of the gone mind?

\vspace{0.2 cm}

One of the most interesting and puzzling components of Asimov's vision of 
a future with robots playing a major role in our society was the existence of a 
novel research field: {\em robopsychology}. Experts in this area had to deal 
with the sometimes unexpected behavior of robots, emerging from the 
conflicts associated to the there rules of robotics and their inevitable interaction 
with a complex external world. In the picture presented here, a different (but related) 
class of robotic psychology might emerge in the future. The interaction, merging and 
blurring of behavioral patterns resulting from the HRI described above defines 
an uncharted territory. Long before machines might 
outsmart us or develop consciousness or intelligent behaviour (Barrat 2013) we will need to 
either face the rise of the humanbot or prevent it to happen. 




\section*{Acknowledgments}
The author would like to thank Luis Seoane, Marti Sanchez-Fibla 
and Paul Vershure for a critical reading of the manuscript and useful discussions on 
artificial intelligence, evolved agents and complexity. Also to Joan Manel Sol\'e, with whom 
I discussed and wrote as freshmen my first codes implementing the ELIZA program. 
This work has been supported by the Bot\'in Foundation by Banco Santander through its
Santander Universities Global Division, a MINECO FIS2015-67616 fellowship 
and by the Santa Fe Institute.

\section{References}

\begin{enumerate}

\item
 Amari, S. I., and Maginu, K. (1988). Statistical neurodynamics of associative memory. 
 Neural Networks 1, 63-73. 
  
\item
Amit, D. J. (1992). {\em Modeling brain function: The world of attractor neural networks}. 
Cambridge University Press.  
  
\item
Arel, I., Rose, D. C., and Karnowski, T. P. 2010. Deep machine learning-a new frontier in artificial intelligence research [research frontier]. Comput. Intell. Magazine IEEE 5, 13-18.

\item
Barrat, J. (2013) {\em Our final invention. Artificial intelligence and the end of the human era.} Thomas 
Dunne Books. New York. 

 \item
 Bradshaw, Jeffrey M., et al. 2004. Making agents acceptable to people. 
 In: {\em  Intelligent Technologies for Information Analysis}. Springer Berlin Heidelberg, pp. 361-406. 
   
 \item
 Breazeal, C. L. 2004. {\em Designing sociable robots}. MIT press.
    
 \item
 Breazeal, C. L. et al. 2016. Social Robotics. In: {\em Springer Handbook of Robotics}. Springer. 
 pp. 1935-1972.
 
\item
Buttazzo, G. 2001. Artificial consciousness: Utopia or real possibility?. Computer 34, 24-30.
  
\item
Clark, A. and Grush, R. (1999). Towards a cognitive robotics. Adaptive Behavior, 7, 5-16.

\item
Edelman, G. 1992. {\em Bright air, brilliant fire}. Basic Books.

\item
Fellous, J. M. and Arbib, M. A. (Eds.). 2005. 
{\em Who needs emotions?: The brain meets the robot}. Oxford University Press.

\item
Floreano, D., Ijspeert, A. J. and Schaal, S. 2014. Robotics and Neuroscience. Curr. Biol. 24, R910-R920.

\item
Gorbenko, A., Popov, V. and Sheka, A. 2012. Robot self-awareness: 
Exploration of internal states. Applied Math. Sci. 6, 675-688.

\item
Holland, O. 2003. {\em Machine consciousness}. Imprint Academic.

\item
Ienca, M., Jotterand, F., Vic?, C. and Elger, B. (2016). Social and Assistive Robotics in Dementia Care: Ethical Recommendations for Research and Practice. International Journal of Social Robotics 8, 565-573.

\item
Kahn P. H., Friedman, B., Perez-Granados, D. R. and Freier, N. G. (2004, April). Robotic pets in the lives of preschool children. Interaction Studies 7, 405-436.

\item
Kidd, C. D., Taggart, W., and Turkle, S. (2006). A sociable robot to encourage social interaction among the elderly. In Proceedings 2006 IEEE International Conference on Robotics and Automation, 2006. ICRA 2006. pp. 3972-3976. 

\item
Koch, C. 2015. Do Androids Dream?. Sci. Am. Mind 26, 24-27.

\item
LeCun, Y., Bengio, Y. and Hinton, G. 2015. Deep learning. Nature, 521, 436-444.

\item
Marques, H. G. and Holland, O. 2009. 
Architectures for functional imagination. Neurocomputing 72, 743-759.

\item
Melson, G. F., Kahn Jr, P. H., Beck, A. and Friedman, B. (2009). Robotic pets in human lives: Implications for the human-animal bond and for human relationships with personified technologies. Journal Social Issues, 65, 545-567.

\item
Mnih, V., Kavukcuoglu, K., Silver, D., Rusu, A. A. et al. 2015. 
Human-level control through deep reinforcement learning. Nature 518, 529-533.

\item
Noda, K., Arie, H., Suga, Y. and Ogata, T. (2014). Multimodal integration learning of robot behavior using deep neural networks. Robotics and Autonomous Systems 62, 721-736.

\item
Nourbakhsh, I. R. 2013. {\em Robot futures}. MIT Press.

\item
Oztop, E., Kawato, M. and Arbib, M. 2006. Mirror neurons and imitation: A computationally guided review. Neural Networks 19, 254-271.

\item
Prescott, T. J., Gonzalez, F. M. M., Gurney, K., Humphries, M. D. and Redgrave, P. (2006). A robot model of the basal ganglia: behavior and intrinsic processing. Neural Networks, 19, 31-61.

\item
Rault, J. L. (2015). Pets in the Digital Age: Live, Robot, or Virtual?. Frontiers Vet. Science, 2.

\item
Reitz, C., Brayne, C. and Mayeux, R. (2011). Epidemiology of Alzheimer disease. Nature Rev. Neurology, 7, 137-152.

\item
Reggia, J. A. 2013. The rise of machine consciousness: Studying consciousness with computational models. 
Neural Networks 44, 112-131.

\item
Rojas, R. (2013). {\em Neural networks: a systematic introduction}. Springer Science.

\item
Sol\' e, R. (2016). Synthetic transitions: towards a new synthesis. Phil. Trans. R. Soc. B, 371, 20150438.

\item
Sporns, O., and Alexander, W. H. (2002). Neuromodulation and plasticity in an autonomous robot. Neural Networks, 15, 761-774.

\item
Sung, J., Grinter, R. E. and Christensen, H. I. (2009, April). Pimp my roomba: Designing for personalization. In Proceedings of the SIGCHI Conference on Human Factors in Computing Systems (pp. 193-196). ACM.

\item
Suddendorf, T. 2013. {\em The gap: The science of what separates us from other animals}. Basic Books.

\item
Takayanagi, K., Kirita, T., and Shibata, T. (2014). Comparison of Verbal and Emotional Responses of Elderly People with Mild/Moderate Dementia and Those with Severe Dementia in Responses to Seal Robot, PARO. 
Frontiers Aging Neurosci. 6, 257. 

\item
Takeno, J. 2012. {\em Creation of a Conscious Robot: Mirror Image Cognition and Self-Awareness}. CRC Press.

\item
Tononi, G. and Koch, C. 2015. Consciousness: here, there and everywhere?. 
Phil. Trans Royal Soc. London B 370, 20140167.

\item
Tononi, G. 2012. $\Phi$, {\em Phi, a voyage from the brain to the soul}. Pantheon Books. New York.

\item
Tononi, G. and Edelman, G. M. 1998. Consciousness and complexity. Science 282, 1846-1851.

\item
Tononi, G., Boly, M., Massimini, M. and Koch, C. (2016). Integrated information theory: from consciousness to its physical substrate. Nature Reviews Neuroscience (in press).

\item
Turing, A. M. 1950. Computing machinery and intelligence. Mind 59, 433-460.

\item
Verschure, P. F., Kröse, B. J. and Pfeifer, R. (1992). Distributed adaptive control: The self-organization of structured behavior. Robotics and Autonomous Systems, 9, 181-196.

\item
Verschure, P. F., Pennartz, C. M. and Pezzulo, G. (2014). The why, what, where, when and how of goal-directed choice: neuronal and computational principles. Phil. Trans. R. Soc. B, 369, 20130483.

\item
Verschure, P. F. (2016). Synthetic consciousness: the distributed adaptive control perspective. 
Phil. Trans. R. Soc. B, 371, 20150448.

\item
Weizenbaum, J. (1966). ELIZA: a computer program for the study of natural language communication between man and machine. Comm. ACM. 36-45.

\item
Zarkadakis, G. (2015). {\em In Our Own Image: Will artificial intelligence save or destroy us?}. 
Random House. New York. 

\end{enumerate}

\bibliographystyle{frontiersinSCNS_ENG_HUMS}

\bibliography{test}

\end{document}